\documentclass[reprint,amsmath,amssymb,aps,showpacs]{revtex4-1}  
\pdfoutput=1

\usepackage{graphicx}
\usepackage{dcolumn}
\usepackage{bm}  
  
\begin{document}

\title{Discontinuous transitions can survive to quenched disorder in a two-dimensional  nonequilibrium system
} 

\author{ Minos A. Neto$^{\ddag}$ and E. Brigatti$^{\star}$
}

\affiliation{$^{\ddag}$ Departamento de F\'{\i}sica, Universidade Federal do Amazonas, 3000, Japiim, 69077-000, Manaus, AM, Brazil}
\affiliation{$^{\star}$ Instituto de F\'{\i}sica, Universidade Federal do Rio de Janeiro, 
Av. Athos da Silveira Ramos, 149,
Cidade Universit\'aria, 21941-972, Rio de Janeiro, RJ, Brazil}
\affiliation{e-mail address: edgardo@if.ufrj.br}

\begin{abstract}

We explore the effects that quenched disorder has on discontinuous nonequilibrium phase transitions into absorbing states. We focus our analysis on the Naming Game model, a nonequilibrium low-dimensional system with 
different absorbing states.
The results obtained by means of the finite-size scaling analysis and from the study of the temporal
dynamics of the density of active sites near the transition point
evidence that the spatial quenched disorder does not destroy the discontinuous transition.


\end{abstract}
 
\maketitle

\section{Introduction}

The study of nonequilibrium models 
is currently an important topic of statistical physics \cite{odor0,odor}.
The theory of phase transitions in equilibrium
systems is well established and rests upon solid foundations,  
with numerous results rigorously proved.
Moreover, it has been thoroughly tested and validated through
a large number of empirical and numerical evidence. 
In contrast, work on the development of a similar theory 
for nonequilibrium systems is rather recent, 
and 
results are often limited in scope and lack adequate validation \cite{odor0,odor}.
For this reason, one is sometimes led to apply most of the fundamental equilibrium
results to nonequilibrium systems
following simple analogies or heuristic generalizations.
The question of whether some specific 
result of equilibrium statistical mechanics can 
be effectively extended to nonequilibrium
systems is open for a number of different issues.
In this paper we will focus on an important result 
of equilibrium statistical physics,
which states that the introduction
of quenched random fields or
interactions in low-dimensional ($d\le2$) systems  
causes the disappearance of discontinuous phase transitions \cite{equilibrium}.
Quenched randomness precludes the presence of such 
types of transitions.





In  a recent work Villa Mart\'in {\it et al.} \cite{villa} addressed the question whether 
such result can be translated to the nonequilibrium realm.
They studied a two-dimensional reaction-diffusion contact-process-like model \cite{odor0,odor}, a very simple nonequilibrium model with one absorbing state 
which exhibits a discontinuous transition.
They showed that 
the introduction of disorder annihilated the discontinuous phase transition and induced a continuous one, just as what should be expected in the equilibrium case. 
This result 
led to the conjecture that the arguments used in equilibrium systems could be extended to nonequilibrium ones, 
leading to the disappearance of the discontinuous phase transitions, due to quenched disorder, 
in general low-dimensional nonequilibrium systems with absorbing states \cite{villa}.

Intrigued by this fascinating idea, we test here this hypothesis
by means of a recently introduced  nonequilibrium model with absorbing states: 
the Naming Game model \cite{baronchelli06}.
This model is related to the large family of models which implement
an ordering dynamics that may generate global consensus as an emergent 
phenomenon among interacting agents.
Despite its simplicity, its collective dynamics present
new features not commonly found in other more traditional models. 
In fact, these dynamics rests on a memory-based negotiation, where trials shape and reshape the memories, allowing for intermediate individual states, feedback phenomena, and dynamic inventories.
This model generated a vivid interest because some of its variants were 
able to describe different aspects of linguistic dynamics, like,
for example, the birth of neologisms \cite{baronchelli06}, 
the effects of reputation on fixing vocabulary \cite{edo1},
the self-organization of a hierarchical category structure \cite{apply1}, 
the emergence of universality in color naming patterns \cite{apply2},
the duality of patterning in human communication \cite{apply3}, and
the rise of protosyntactic structures \cite{apply4}.

Some versions of the model display a discontinuous phase transition \cite{baronchelli07,eddy}.
This behavior can be obtained by the introduction of a specific control parameter which represents the efficiency of the communication process, 
accounting for external or internal influences  
or agents' irresolute attitude \cite{baronchelli07}.
This ingredient generates a transition between an absorbing state of 
global consensus and a stationary state with several coexisting conventions.
The richness of  
the Naming Game  model, which  moves a step further from the simplicity of the contact process, 
does not preclude the possibility of describing it
with methods borrowed from classical equilibrium statistical mechanics.
This fact could be appreciated in previous studies which clearly characterized its phase transition \cite{baronchelli07,nos}.

\section{The model}

We simulate the game on a regular two-dimensional (2D) square lattice with periodic boundary conditions,
where only nearest-neighbor pairs of sites are
considered in the rules of the game. 
At each  of the  $L\times L$ sites an agent, or player, is placed. It is 
characterized by an inventory which can store an infinite number of conventions.
This inventory is structured as an array of potentially infinite 
cells, 
where each cell is set on one of a countable infinite number of possible states.
Every player starts with an empty inventory.
At each time step, a pair of agents is randomly selected.
The first agent 
selects one of its conventions 
or creates a new one, if its inventory is empty. 
After that, the convention selected  
is transmitted to the second agent. 
If this last agent already possesses in its own inventory the convention transmitted,
the two agents involved in the interaction update their inventories 
so as to keep only the considered convention, with a probability $\beta$.  
Conversely, no action is 
performed by the couple of agents, with probability $1-\beta$.
Otherwise, if the second agent does not already possess the transmitted convention,
the interaction is a failure, and it adds the new convention to its own inventory.
If $\beta=1$, the model coincides with the original Naming Game \cite{baronchelli06}
implemented on a low-dimensional lattice. 
This model always converges to an absorbent ordered state characterized
by one single convention adopted by every agent \cite{baronchelli06b}. 
When the behavior of the model is studied as a function of the parameter $\beta$, a transition
between overall consensus and several coexisting conventions appears, driven by the  
value of the parameter.
The critical value $\beta_c$ is close to 1/3
for the mean-field model \cite{baronchelli07}
and $\beta_c=0.329 \pm 0.001$
on a regular 2D lattice \cite{nos}.
Moreover, this last study shows,
by means of a rigorous finite-size scaling analysis,
that the model effectively displays a discontinuous 
nonequilibrium phase transition with absorbing states \cite{nos}.

This pure model can be modified to obtain a disordered version. In this version, the parameter that controls the outcome of each interaction is no longer the same for all agents. Each agent $i$ is now characterized by a random uncorrelated probability $\beta_i$, where we take $\beta_i=\beta+r$ where $\beta$ is constant and $r$ is a random number on the interval  $[-\eta,+\eta]$ chosen 
with a uniform probability.
The values of $\beta_i$ are thus randomly defined and fixed at the beginning of each simulation.

\section{Results and discussion}

\begin{figure}[h]
\begin{center}
\vspace{0.6cm}
\includegraphics[width=0.5\textwidth, angle=0]{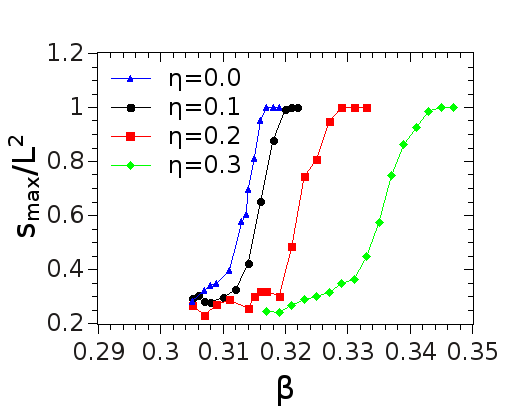}
\end{center}
\caption{\small Characterization of the phase transition 
using 
the mean of the normalized largest cluster
size as a function of $\beta$. Results are displayed for different 
levels of noise ($\eta$). 
Each point is averaged over $100$ simulations, $L=40$.}
\label{fig_fragments}
\end{figure}


As a first step, we analyze the differences generated 
in the general form of the phase transition 
by the introduction of the quenched noise. 
The phase transition corresponds to the shift from an active stationary state, characterized by disordered and fragmented clusters, to an absorbing state made of a single cluster represented by the same convention.
Because of this phenomenology, the relative size of the largest cluster
present in the system is an excellent parameter that characterizes
the transition \cite{castellano00,brigatti15}.
This parameter is defined as the size of the largest cluster, 
made up by the agents sharing the same unique convention,  
normalized by the system size: $s_{max}/L^{2}$. 
This parameter is estimated once the steady state is reached  
and it is averaged over different simulations: $\langle s_{max}/L^{2}\rangle$.
As the system presents very slow relaxation time close to the transition,
we have to run  very long simulations  
($4 \times 10^{10}$  Monte Carlo steps) 
which forced us to adopt $L$ values only up to 100. 
In Figure \ref{fig_fragments} we can observe 
how the introduction of noise, controlled by the parameter $\eta$,
impacts the transition behavior.
The critical value of $\beta_{c}(L)$ is shifted towards higher $\beta$
values but, surprisingly, the discontinuous phase transitions are not clearly rounded 
by disorder, a fact that generates some doubts 
about the potentiality of noise in generating a continuous transition. 


\begin{figure}[h]
\begin{center}
\vspace{0.6cm}
\includegraphics[width=0.3\textwidth, angle=0]{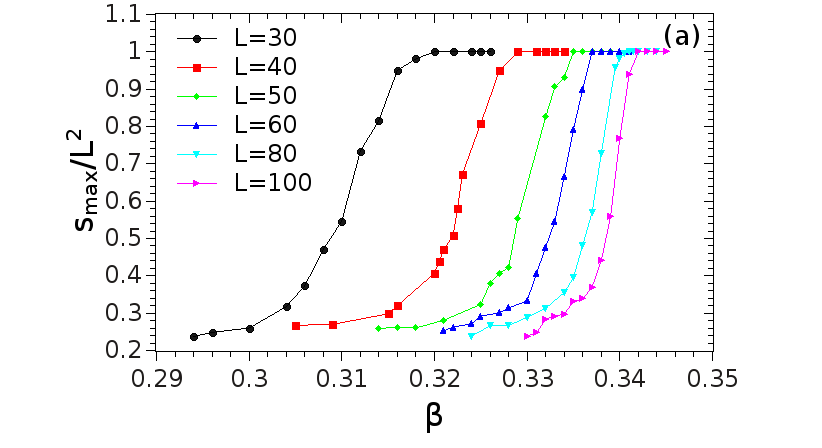}
\includegraphics[width=0.3\textwidth, angle=0]{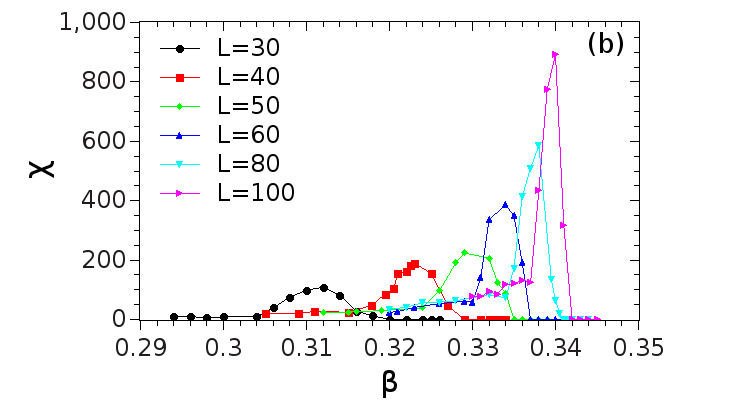}
\includegraphics[width=0.3\textwidth, angle=0]{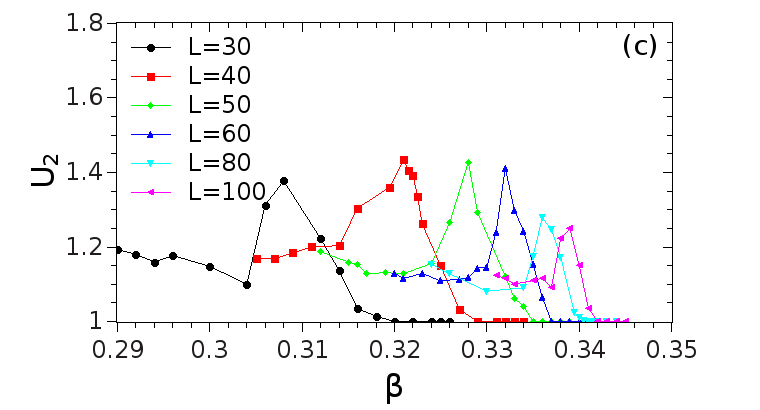}
\end{center}
\caption{\small From top to bottom: (a) Mean of the normalized largest cluster
size, (b) 
its variance $\chi$ and (c) the moment ratio $U_2$
as a function of $\beta$ for different system sizes
and $\eta=0.2$. 
Each point is averaged over up to $200$ simulations.}
\label{fig_phtrans}
\end{figure}

For this reason we turn our attention to
a rigorous characterization of 
the phase transition,
based on a finite-size scaling analysis.
In fact, the scaling behavior of the system
near the transition can clearly discriminate between a 
continuous or discontinuous transition.

We performed the analysis evaluating the fluctuations of the size of the largest cluster:
\begin{eqnarray}
\chi=L^{2}(\langle s_{max}^{2}\rangle - \langle s_{max} \rangle^{2}) 
\nonumber
\end{eqnarray}
and its moment ratio (reduced cumulant) \cite{dickman98}:
\begin{eqnarray}
U_2=\frac{\langle s_{max}^{2} \rangle}{\langle s_{max} \rangle^{2}};
\nonumber
\end{eqnarray}
for different $L$ values.

\begin{figure}[t]
\begin{center}
\includegraphics[width=0.22\textwidth, angle=0]{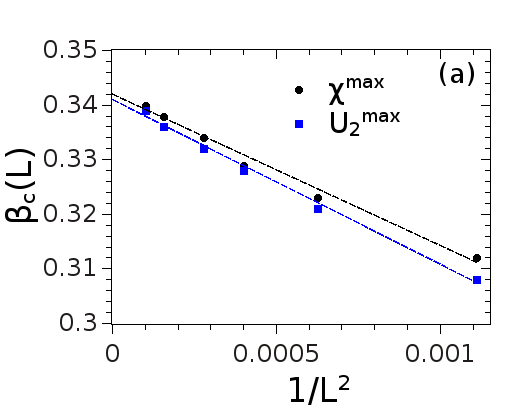}
\includegraphics[width=0.22\textwidth, angle=0]{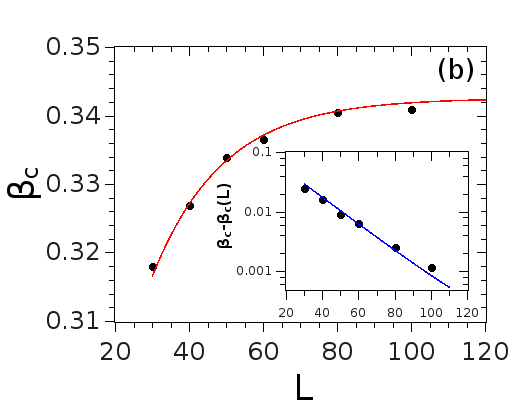}
\includegraphics[width=0.22\textwidth, angle=0]{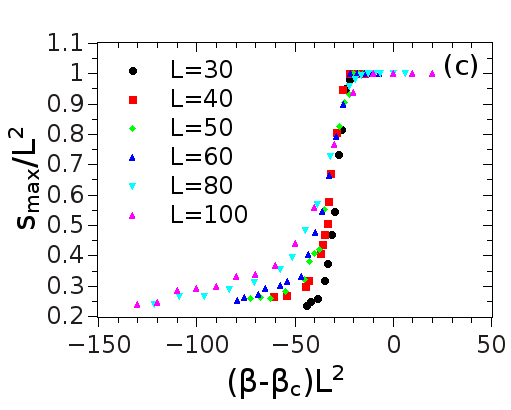}
\includegraphics[width=0.22\textwidth, angle=0]{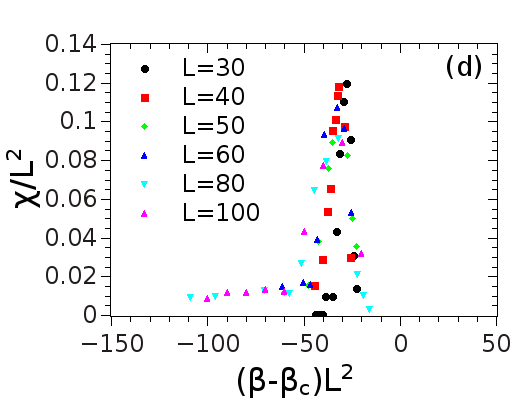}
\end{center}
\caption{\small 
{\it Top}: 
(a) Convergence to the asymptotic transition point $\beta_c$ 
of the finite-size transition points $\beta_c(L)$ measured from
the variance and the $U_2$ maxima.
(b) Convergence to the asymptotic transition point $\beta_c$ 
of the finite-size transition points $\beta_c(L)$ measured from the
jump location in the normalized size of the largest cluster. 
The continuous line represents the best fitting function:
$\beta_c(L)=0.343-0.125 \exp(-0.052 L)$.
In the inset, the semilogarithmic plot shows in detail the expected exponential behavior of the difference $\beta_c-\beta_c(L)$.
\textit{Bottom}: 
(c) Rescaled plot for $\langle s_{max} \rangle/L^2$ 
and (d) its fluctuations. 
}
\label{fig_critic}
\end{figure}

As can be appreciated in Figure \ref{fig_phtrans}, these quantities peak  at around 
$\beta_c(L)$. 
Actually, the use of the maxima of these quantities has proven to be 
a very efficient 
method for performing finite size scaling analysis of discontinuous phase transition into absorbing states \cite{oliveira2015}.
In this case, the asymptotic transition point 
can be obtained by looking at the convergence of the finite size transition points $\beta_c(L)$, 
as estimated by the localization 
of the maxima of the fluctuations or the maxima of the moment ratio.
In both cases, the convergence  is expected to follow an algebraic behavior:
$  \beta_c(L)=\beta_c+aL^{-2}$, 
which is the usual equilibrium scaling for discontinuous transitions \cite{oliveira2015,mio19,privman}.
Our data follow  these scaling laws very well:
Figure \ref{fig_critic} 
shows how the maxima positions
for $\chi$ and $U_2$ effectively decrease as $1/L^2$. 
An extrapolation for $L\to \infty$ yields $\beta_c= 0.342\pm0.001$ 
for $\chi$ and   $\beta_c= 0.341\pm0.001$ for $U_2$,
 two very close values.

An alternative approach \cite{borgs} for the estimation of the 
asymptotic transition point uses 
the location of the observed discontinuity 
in  the normalized size of the largest cluster
(the first value of $\beta$ for which $\langle s_{max} \rangle/L^2$
is smaller than $1$).
Using this method the convergence is supposed to be exponential
and the extrapolation for $L\to \infty$ gives $\beta_c= 0.343\pm0.002$ (see
Figure \ref{fig_critic}).  
In the same figure we can observe how the difference $\beta_c-\beta_c(L)$ 
clearly presents the expected exponential behavior. 

Additional consistency checks of the above results can be performed
verifying whether  the measured quantities present  
the typical scaling of a discontinuous transition near the 
transition point, a standard procedure for equilibrium finite-size scaling analysis \cite{binder,oliveira2015}.
The scaling plot of $\langle s_{max} \rangle/L^2$     
should be obtained 
introducing the rescaled control parameter $\beta^*=  (\beta-\beta_c)L^{d}$,
where $d$ is the system dimension.
Similarly, the scaling plot of the fluctuations should be drawn
using the rescaled fluctuations $\chi \cdot L^{-d}$ and the rescaled  parameter $\beta^*$.
Figure \ref{fig_critic} shows
a reasonable collapse which satisfies these relations, 
strongly suggesting the validity of the finite-size 
scaling ansatz expected for a discontinuous transition.
\\

An alternative way for differentiating between continuous and discontinuous transition
is based on the distinct behavior of the temporal dynamics of the density of active sites. 
In our model such density corresponds to the  fraction $\rho$ of agents 
presenting more than one convention in their inventories \cite{lipowska}.
Near the transition point  the density is expected to decay 
smoothly
to zero in the case of  continuous phase transitions 
and to develop an abrupt discontinuous jump in the case of discontinuous transitions.
In particular, in continuous phase transitions the density of active sites typically presents 
a power-law decay \cite{odor}. Even so, some models which 
present continuous transitions between active and absorbing states 
are characterized by slower decays (logarithmic) of active sites (for example, in voter-like models \cite{22}   and in Potts model with absorbing states \cite{23}).

\begin{figure}[t]
\begin{center}
\includegraphics[width=0.5\textwidth, angle=0]{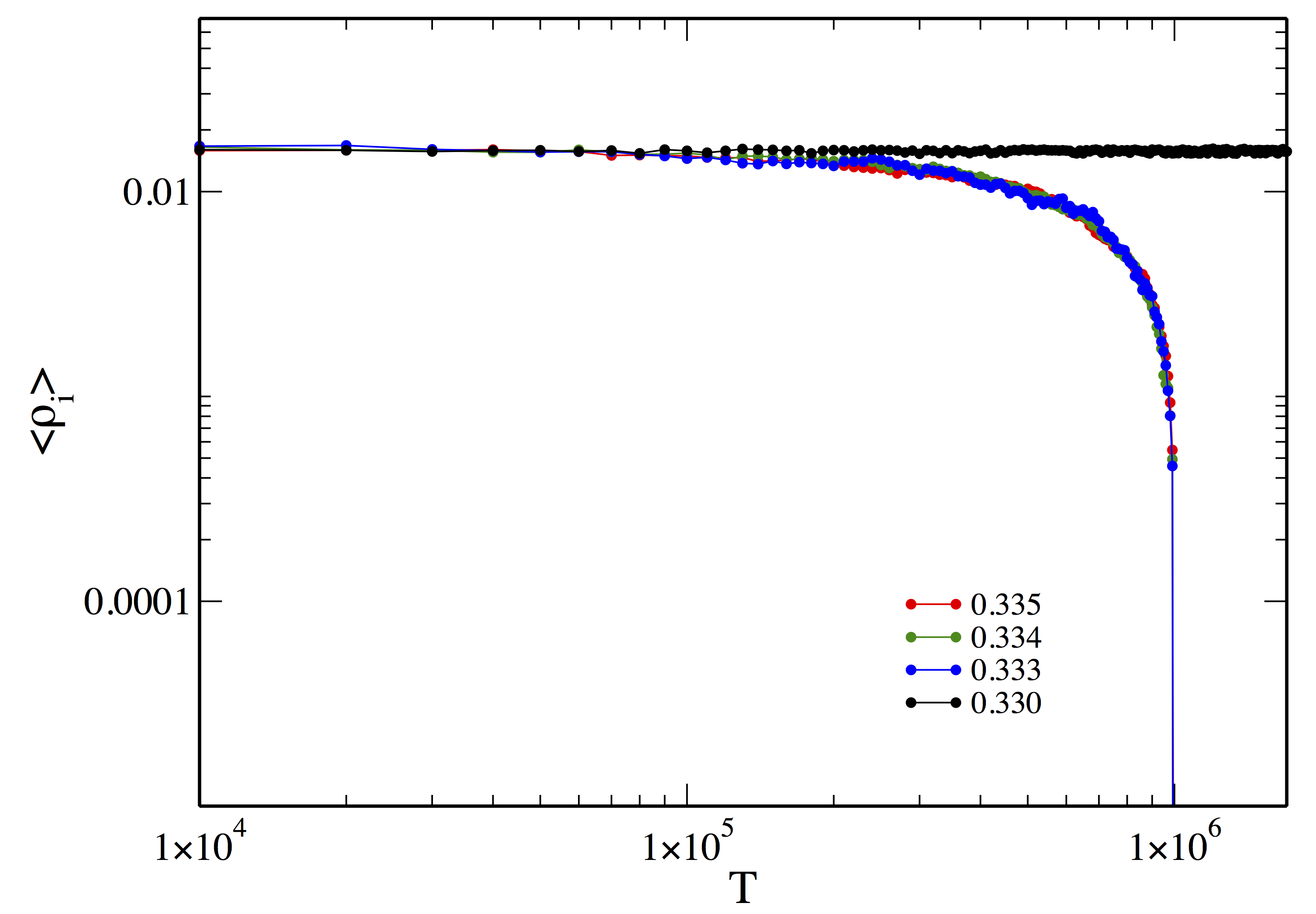}
\end{center}
\caption{\small 
Time evolution of the density of multi-convention agents for
different values of $\beta$, $\eta=0.2$,  and $L=60$.
The rescaled time $T=(t-t^0_i)+10^6$, 
where $t$ is the Monte Carlo step and $t^0_i$ corresponds to the step when 
the density $\rho_i$  goes to zero. 
Points represent the ensemble average of $\rho_i$,
obtained considering only the simulations
which converged to the consensus state.
For $\beta=0.330$, 
the time is not rescaled and the average is over all the simulations.
}
\label{fig_density}
\end{figure}

We collected some numerical data depicting the dynamics of the density of active sites
for our model with quenched disorder. 
A clear representation of the density evolution
can be obtained 
averaging over many independent runs. 
Before calculating these averages, since the convergence time in this model is characterized
by a very  wide distribution, it is useful to shift each time series $i$ by the time $t^0_i$, which corresponds to the time when the density $\rho_i$  goes to zero.
As can be seen in Figure \ref{fig_density}, the time evolution 
of the density of active sites clearly develops an abrupt discontinuous jump,
corroborating the view that the transition is discontinuous.\\

In summary, we have numerically studied 
the effects of introducing a spatial quenched disorder in a
Naming Game model which, in its pure version, presents a well known
discontinuous phase transition.
The aim of the analysis is to state whether the presence of the quenched randomness  results in the disappearance of the discontinuous phase transition, as is known to happen in 
equilibrium systems.
In contrast to this conjecture, 
 our analysis 
 suggests that the model maintains the discontinuous
phase transition.
This conclusion is supported 
by the evidence provided by the results 
of a finite-size scaling analysis. 
We found that, as for the pure (without disorder) model, the 
behavior of the finite-size transition points
 measured from the variance, the moment ratio or, the jump locations
show a scaling behavior which can be associated with
a  discontinuous transition.  
Additional confirmations of these results come from 
the collapse of the scaling plot of 
the order parameter 
and its fluctuations which have been obtained using the scaling 
law expected for a discontinuous transition \cite{oliveira2015}. 
Moreover, 
the temporal dynamics of the density of the active sites
near the transition point 
 develops an abrupt discontinuous jump, 
 as expected in the case of discontinuous transitions. 
These numerical results  constitute evidence for
the 
survival of the discontinuous phase transition
after the introduction of quenched disorder.

Previous works have already shown that temporal disorder does not destroy 
 discontinuous transitions in  a variety of two-dimensional models \cite{fiore}.
 In our study, the transition surprisingly survives also in the case of a simple 
 quenched spatial  disorder.
 It is interesting to note that the  model considered in \cite{villa}, for which
 the discontinuous transition disappeared, is characterized by only one absorbing
 state. In contrast, our model  presents  a large number of possible different 
 absorbing states.

We  conclude our work by proposing a heuristic argument
for explaining the phenomenology behind this single-multiple
absorbing states dichotomy in rounding (or not) the discontinuous transition. 
An interesting inspiration comes from the case of continuous phase transitions.
In that scenario 
the result that random fields destroy an equilibrium phase 
transition in 
low dimensions \cite{ma}
cannot be transposed to nonequilibrium systems. 
This fact was shown by the analysis of Barghathi {\it et al.}, where the phase transition of 
a generalized contact process presenting 
two absorbing states
persists in the presence of 
disorder \cite{barga}. 
In such an example the existence of 
two absorbing states is essential.
In fact, along its dynamics the system organizes
in distinct uniform domains corresponding to the two absorbing states 
and neither active sites nor new domains 
can arise in the interior of a given 
domain, as its states are inactive.
After a process of coalescence of these domains, in 
the long-time limit, the system reaches a single-domain state.
In contrast, in the equilibrium case of a random-field Ising model, 
the growth of a uniform domain is limited by 
spin flips which can occur anywhere due to fluctuations. 
The domains size reaches a typical value controlled by 
the Imry-Ma argument \cite{ma}, suppressing the continuous
transition. 
The case of the discontinuous transition can be explained 
following the work of 
Kardar {\it et al.} \cite{kardar} which suggests that disorder precludes
discontinuous transitions by generating 
islands of arbitrary size of one of the phases
within the other.
This nested structure of islands  within islands
leads to the formation of hybrid states and
two distinct phases cannot coexist.
This phenomenon can happen in the contact-process-like model considered in \cite{villa}, 
where the unique absorbing state corresponding to no-particle occupation
can pop-up everywhere in the system, as particles disappear at 
random locations. 
In contrast, in our model, once a domain with a specific 
inactive state is defined, it is not possible to create smaller regions 
in the opposite phase inside it, paralyzing this mechanism
and allowing the discontinuous transition to take place.



\section*{Acknowledgments} 

M.A.N. acknowledges support from FAPEAM and  Prof. Jos\'e Luiz de Souza Pio, from Laborat\'orio de Rob\'otica do IComp/UFAM, for supporting with computing facilities. 
E.B. thanks FAPEAM and  the Departamento de F\'{\i}sica of UFAM for their partial support and hospitality during the realisation of part of this work.



\end{document}